\documentclass[aps,prl,twocolumn,groupedaddress,showpacs]{revtex4}
\usepackage{graphicx}
\usepackage{dcolumn}
\usepackage{bm}
\usepackage{amsmath,amsfonts,amssymb,amsthm,verbatim}
\usepackage[ansinew]{inputenc}
\usepackage[usenames]{pstcol}

\usepackage{color}
\usepackage{epsfig}

\newcommand{\be}{\begin{equation}}
\newcommand{\ee}{\end{equation}}
\newcommand{\bea}{\begin{eqnarray}}
\newcommand{\eea}{\end{eqnarray}}
\newcommand{\beq}{\begin{eqnarray}}
\newcommand{\eeq}{\end{eqnarray}}

\newcommand{\bi}{\begin{itemize}}
\newcommand{\ei}{\end{itemize}}
\newcommand{\bc}{\begin{center}}
\newcommand{\ec}{\end{center}}

\begin{document}

\title{Synthetic helical liquid in a quantum wire}
\author{George I. Japaridze$^{1,2}$, Henrik Johannesson$^3$, and Mariana Malard$^4$}
\affiliation{$\mbox{}^1$Andronikashvili Institute of Physics,
Tamarashvili 6, 0177 Tbilisi, Georgia} \affiliation{$\mbox{}^2$Ilia
State University, Cholokasvili Avenue 3-5, 0162 Tbilisi, Georgia}
\affiliation{$\mbox{}^3$Department of Physics, University of
Gothenburg, SE 412 96 Gothenburg, Sweden}
\affiliation{$\mbox{}^4$Faculdade UnB Planaltina, University of
Brasilia, 73300-000 Planaltina-DF, Brazil}

\begin{abstract}

We show that the combination of a Dresselhaus interaction and a
spatially periodic Rashba interaction leads to the
formation of a helical liquid in a quantum wire when the
electron-electron interaction is weakly screened.
The effect is sustained by a helicity-dependent effective band gap
which depends on the size of the
Dresselhaus and Rashba spin-orbit couplings. We
propose a design for a semiconductor device in which the helical
liquid can be realized and probed experimentally.

\end{abstract}


\maketitle

The concept of a {\em helical liquid} $-$ a phase of matter where
spin and momentum directions of electrons are locked to each
other $-$ underpins many of the fascinating features of the recently
discovered topological insulators \cite{Review}. In the case of an
ideal two-dimensional (2D) topological insulator, electron states at
its edges propagate in opposite directions with opposite spins,
forming a one-dimensional (1D) helical liquid (HL)
\cite{KaneMele,BHZ}. Given the right conditions \cite{Wu,Xu}, the
spin-filtered modes of the HL may serve as ballistic
conduction channels \cite{Konig}, holding promise for novel
electronics/spintronics applications.

The HL is expected to exhibit several unusual properties, such as
charge fractionalization near a ferromagnetic domain wall \cite{Qi},
interaction-dependent  response to pinching the sample into a point
contact \cite{SJ}, and enhanced
superconducting correlations when two HLs are coupled together
\cite{Tanaka}. A particularly tantalizing scenario is the
appearance of Majorana zero modes in an HL in proximity to a
superconductor and a ferromagnet \cite{FuKane}. However, testing
these various predictions in experiments on the HgTe/CdTe
quantum well structures in which the HL phase has been observed
is a formidable challenge: The softness and reactivity of HgTe/CdTe makes it
difficult to handle \cite{Molenkamp}, and moreover, charge puddles formed due
to fluctuations in the donor density may  introduce a helical edge resistance
\cite{Vayrynen}.
Alternative realizations of the HL are therefore in high
demand. The prospect of using the dissipationless current of an HL in
future chip designs adds to the importance of this endeavor \cite{IBM}.
\begin{figure}[htbp] 
   \centering
   \includegraphics[width=2.4in]{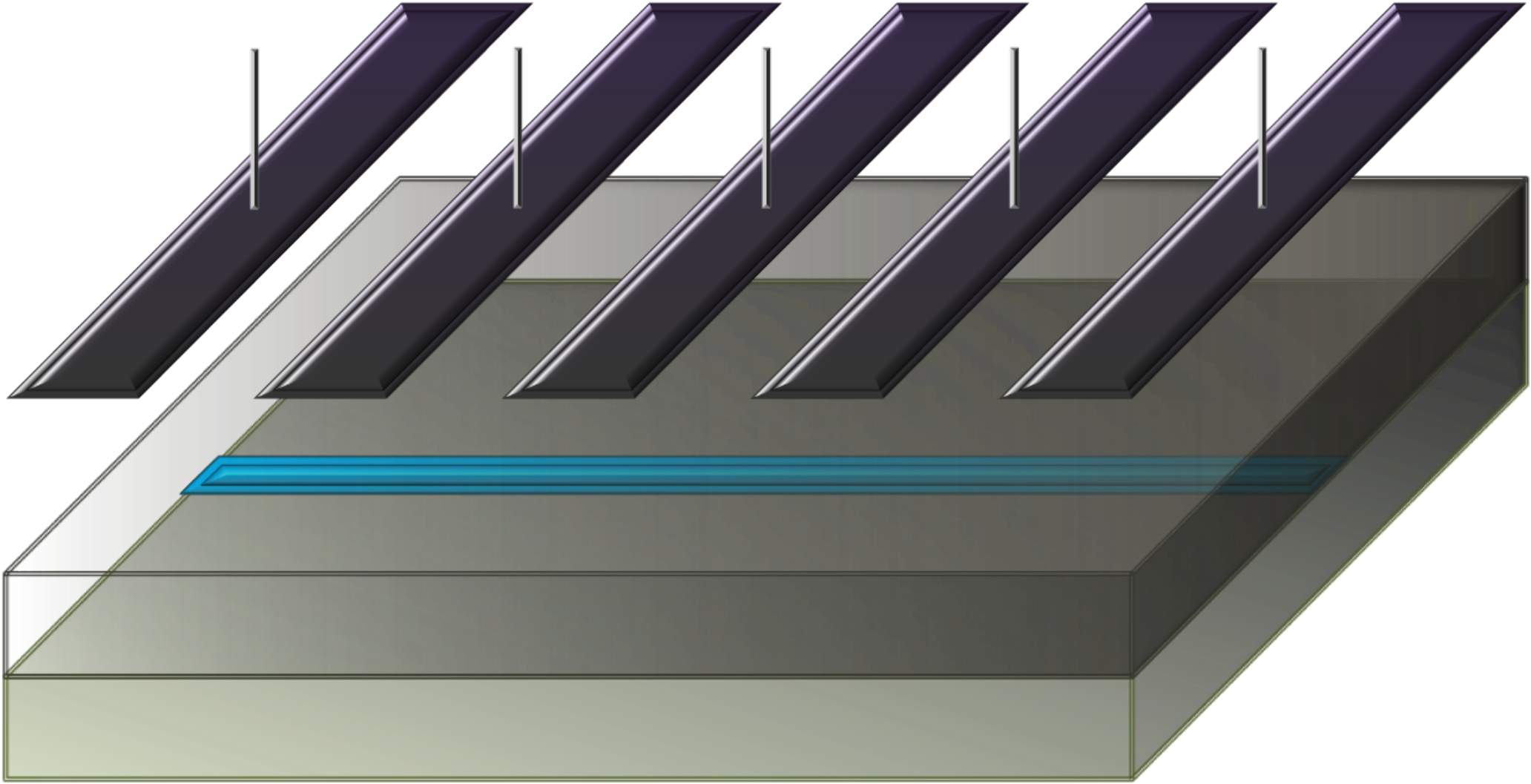}
   \caption{Device supporting a 1D synthetic helical liquid:
   Electrons in a single-channel quantum wire (blue) formed in a heterostructure
   supporting a Dresselhaus interaction are
   subject to a modulated Rashba field from a periodic sequence of charged
   top gates (dark grey).}
   \label{fig:example}
\end{figure}

One suggestion is to use a nanowire made of a ``strong
topological insulator'' material \cite{Review}. When pierced with a
magnetic flux quantum, the electrons in the wire
are predicted to form an interacting HL \cite{Egger}. In another
scheme $-$ appearing in
attempts to engineer Kitaev's toy model \cite{Kitaev} for p-wave
pairing \cite{LutchynOreg} $-$ electrons in a quantum wire form an
HL when subject to a Rashba spin-orbit coupling combined with a
transverse magnetic field \cite{StredaSeba}. These, like most other
proposals for HLs in quantum wires
\cite{Braunecker}, specifically rely on the presence of a magnetic
field.

In this article we show that an HL can be produced and controlled in a
quantum wire using electric fields only. The advantages of employing electric
fields rather than magnetic fields are manifold. Most importantly, an electric field
does not corrupt the feature that counterpropagating helical modes carry antiparallel
spins. Also, an electric field can easily be
generated and applied locally, and eliminates many of the design complexities
that come with the use of magnetic fields \cite{AwschalomSamarth}. Our proposed device
(see FIG. 1)  exploits
an unexpected effect that appears when interacting electrons are subject
to a Dresselhaus spin-orbit interaction combined with a {\em spatially
periodic} Rashba interaction:
When the electron density is tuned to a certain value,
determined by the wavelength of the Rashba modulation,
a band gap tied to the helicity of the electrons opens. This gives rise to an HL.
Notably, the required setup for realizing this HL is built around standard
nanoscale semiconductor technology, and is very different from the recently proposed
all-electric setup in Ref. \onlinecite{Klinovaja} using  carbon nanotubes.
In what follows we derive an effective model that captures the surprising effect
from the interplay between the Dresselhaus and the modulated Rashba interaction. We
analyze the model and explain how the HL materializes, and also discuss the
practicality and robustness of this novel type of a synthetic HL.

We consider a setup with a single-channel quantum wire
formed in a gated 2D quantum well supported by a semiconductor heterostructure.
The electrons in the well are subject to two types of
spin-orbit interactions, the {\em Dresselhaus}  and {\em Rashba}
interactions \cite{Winkler}.  For a heterostructure grown along
$[001]$, with the electrons confined to the
$xy$-plane, the leading term in the Dresselhaus interaction takes
the form $H_\text{D} = \beta (k_x \sigma_x - k_y \sigma_y)$ with
$\beta$ a material-specific parameter.
The Rashba interaction is given by $H_\text{R}
= \alpha (k_x \sigma_y - k_y \sigma_x),$  where $\alpha$
depends on several distinct features of the heterostructure
\cite{Sherman,GolubIvchenko},
including the applied gate electric field. The latter feature
allows for a gate control of the Rashba coupling
$\alpha$  \cite{LiangGao}. It is important to
mention that large fluctuations of $\alpha$ \cite{Sherman}
may drive the HL to an insulating state through an Anderson-type transition \cite{SJJ}.
We shall return to this issue below.

Taking the $x$-axis
along the wire, adding to $H_{\text{D}}$ and $H_{\text{R}}$
the kinetic energy of the electrons as well as the chemical potential,
one obtains $-$ using a tight-binding formulation $-$ the Hamiltonian $H_0 + H_{\text{DR}}$,
where
\begin{eqnarray} \label{uni}
H_0  &\! = \!&\! -t \sum_{n,\alpha}
c^{\dagger}_{n,\alpha}c^{\phantom{\dagger}}_{n+1,\alpha} +
\frac{\mu}{2}\sum_{n,
\alpha}c^{\dagger}_{n,\alpha}c^{\phantom{\dagger}}_{n,\alpha} + \mbox{h.c.}, \\
 H_{\text{DR}} &\! = \!& -\!i\!\sum_{n,\alpha,\beta}
c^{\dag}_{n,\alpha}\!\left[\gamma_{\text{D}}\, \sigma^{x}_{\alpha\beta}
\!+ \!\gamma_{\text{R}}\, \sigma^{y}_{\alpha\beta}\right]
\!c^{\phantom{\dag}}_{n\!+\!1,\beta} \!+\! \mbox{h.c.},
\end{eqnarray}
with $H_{\text{DR}}$ the second-quantized projection of $H_{\text{D}} + H_{\text{R}}$ along the wire.
Here $c^{\dagger}_{n,\alpha}$ ($c^{\phantom{\dag}}_{n,\alpha}$) is
the creation (annihilation) operator for an electron with spin
${\alpha}=\uparrow,\downarrow$ on site $n$ (with spin
along the growth direction $\hat{z}$), $t$ is
the electron hopping amplitude, and $\mu$ a chemical potential
controllable by a back gate. The signs and
magnitudes of $\gamma_{\text{D}}  \!\equiv\! \beta a^{-1}$ and $\gamma_{\text{R}}
\!\equiv \!\alpha a^{-1}$ ($a$ being the lattice spacing) depend on
the material as well as on the particular design of the
heterostructure.

We now envision that we place a sequence of equally charged nanoscale
electrodes on top of the heterostructure (cf. FIG. 1). As a result, the Rashba coupling
will pick up a modulated contribution due to the modulation
of the electric field from the electrodes. Taking their separation
to be the same as their extension along the wire (cf. FIG. 1), we model
the Rashba modulation by a simple harmonic,
\begin{equation} \label{mod}
H_\text{R}^\text{mod}=-i {\gamma_{\text{R}}^{\prime}} \sum_{n,\alpha,\beta}
\cos(Qna) c^{\dag}_{n,\alpha}\sigma^{y}_{\alpha\beta}c_{n+1,\beta}
\,+\,  \mbox{h.c.},
\end{equation}
with $\gamma_{\text{R}}^{\prime}$ the amplitude and
$Q$ its wave number. Besides the modulation of the Rashba interaction, also the chemical potential gets modulated by the external gates: 
\begin{equation} \label{modchempot}
H_\text{cp}^\text{mod}=\frac{\mu^{\prime}}{2} \sum_{n,\alpha}
\cos(Qna) c^{\dag}_{n,\alpha}c_{n,\alpha}
\,+\,  \mbox{h.c.}
\end{equation}
As follows from the analysis in Ref. \onlinecite{Malard}, this term has no effect at low energies unless
the electron density is tuned to satisfy the commensurability condition $|Q - 2k_{F}|<<O(1/a)\,\mbox{mod}\, 2\pi$, with $k_{F}$ the Fermi wave number: At all other densities, including those for which an HL emerges, $H_\text{cp}^\text{mod}$ in Eq. (\ref{modchempot}) is rapidly oscillating and gives no contribution in the low-energy continuum limit. Hence, we shall neglect it here.

Given the full Hamiltonian
$H \!=\! H_0 \!+ \!H_{\text{DR}} \!+ \!H_\text{R}^\text{mod}$,
we pass to a basis which diagonalizes $H_0 \!+ \!H_{\text{DR}}$ in spin space,
\begin{equation} \label{spinor}
 \left( \begin{array}{c}
d_{n,+} \\
d_{n,-} \end{array} \right)
\equiv \frac{1}{\sqrt{2}} \big( \begin{array}{c}
-i e^{-i\theta}c_{n,\uparrow}\,  + \, e^{i\theta}c_{n,\downarrow} \\
 \ \ e^{-i\theta}c_{n,\uparrow} -i e^{i\theta}c_{n,\downarrow}  \end{array} \big),
\end{equation}
with $2\theta=\arctan{\gamma_{\text{D}}/\gamma_{\text{R}}}$.  The index
$\tau\! = \!\pm$ of the operator $d_{n, \tau}$ label the new quantized spin
projections along the direction of the combined Dresselhaus
($\propto \gamma_{\text{D}} \hat{x}$) and uniform Rashba ($\propto \gamma_{\text{R}} \hat{y}$)
fields.
Putting $\gamma_{\text{R}}^{\prime}=0$ in Eq. (\ref{mod}) and
using (\ref{spinor}), the system is found to exhibit
four Fermi points $\pm k_F + \tau q_0$, where $q_{0}a=\arctan\sqrt{(\tilde{t}/t)^{2}-1}$ with
$\tilde{t}=\sqrt{t^{2}+\gamma_{\text{R}}^{2}+\gamma^{2}_{\text{D}}}$, and where
 $k_{F}= \pi \nu/a$  with $\nu=N_{e}/2N$, $N_e \, [N]$ being
 the number of electrons [lattice sites]. The corresponding Fermi
energy $\epsilon_F$ is given by $\epsilon_F =
-2\tilde{t}\cos(k_{F}a)+\mu$.

To analyze what happens when $\gamma_{\text{R}}^{\prime}$ is switched on, we focus on the
physically relevant limit of low energies,
linearize the spectrum around the Fermi points and take the continuum limit $na \rightarrow x$.
By decomposing $d^{\phantom{\dagger}}_{n,\tau}$ into right- and left-moving fields
$R^{\phantom{\dagger}}_{\tau}(x)$ and
$L^{\phantom{\dagger}}_{\tau}(x)$,
\begin{displaymath}  \label{Decomposition}
d^{\phantom{\dagger}}_{n,\tau}  \rightarrow  \sqrt{a}
\big(\mbox{e}^{i(k_F+\tau q_0)x}
R^{\phantom{\dagger}}_{\tau}(x) + \mbox{e}^{i(-k_F +\tau q_0)x} L^{\phantom{\dagger}}_{\tau}(x)\big),
\end{displaymath}
and choosing $|Q - 2(k_F + \tau q_0)| << O(1/a)\,\mbox{mod}\, 2\pi$
one thus obtains an effective theory with two independent branches,
$H\!\rightarrow \sum_{i=1,2}\int dx \ {\cal H}_{i}$, where ${\cal
H}_{1}$ applies to the Fermi points $\pm k_F \mp q_0$, and ${\cal
H}_{2}$ to $\pm k_F \pm q_0$. We here choose $Q=2(k_F + q_0)$, and
come back to the general case below. Omitting all rapidly
oscillating terms that vanish upon integration, one finds
\begin{eqnarray}   \label{linear-Hamiltonian}
{\cal H}_{1}&\!=\!& -iv_F
(:\!R^{\dag}_{-}\partial_{x}R^{\phantom{\dagger}}_{-}\!: -
:\!L^{\dag}_{+}\partial_{x}L^{\phantom{\dagger}}_{+}\!:) \\
{\cal H}_{2}&\!=\!& -iv_F (
:\!R^{\dag}_{+}\partial_{x}R^{\phantom{\dagger}}_{+}\!:-
:\!L^{\dag}_{-}\partial_{x}L^{\phantom{\dagger}}_{-}\!:) \nonumber \\
\label{2linear}
&+&i\lambda\, (\!R^{\dag}_{+}\partial_{x}L^{\phantom{\dagger}}_{-}\!+
\!L^{\dag}_{-}\partial_{x}R^{\phantom{\dagger}}_{+}\! ),
\end{eqnarray}
where $v_{F}=2a\tilde{t}\sin(\pi\nu)$, $\lambda =
 a \gamma^{\prime}_{R}\gamma_{D}(\gamma_{R}^{2}+\gamma^{2}_{D})^{-1/2}$, $:...:$
 denotes normal ordering, and
where we have absorbed the constant phase $\mbox{e}^{i(k_F+q_0)a}$ into $R_+(x)$. 

While the nondiagonal term in Eq. (\ref{2linear}) is renormalization-group (RG) irrelevant in the absence of
{\em e-e} interactions it may turn relevant and open a gap at the Fermi points $\pm k_F \pm q_0$ when the
{\em e-e} interaction
\begin{equation} \label{ee}
H_{e{\text -}e} = \sum_{n,n'; \alpha, \beta} V(n-n')c^{\dagger}_{n,\alpha} c^{\dagger}_{n',\beta}
c^{\phantom{\dagger}}_{n', \beta} c^{\phantom{\dagger}}_{n, \alpha},
\end{equation}
is included. Its low-energy limit can be extracted by following the procedure
from above, and we obtain $H_{e{\text -}e} \!\rightarrow \int dx \ {\cal H}_{e{\text -}e}$, where
\begin{eqnarray}
{\cal H}_{e{\text -}e} & = & g_1
\!:\!R^{\dag}_{\tau}L^{\phantom{\dagger}}_{\tau}L^{\dag}_{\tau'}
R^{\phantom{\dagger}}_{\tau'}\!: +  \, {g}_{2}
\!:\!R^{\dag}_{\tau}R^{\phantom{\dagger}}_{\tau}
L^{\dag}_{\tau'}L^{\phantom{\dagger}}_{\tau'}\!: \nonumber  \\
&+&  \frac{g_{2}}{2}(
:\!L^{\dag}_{\tau}L^{\phantom{\dagger}}_{\tau}
L^{\dag}_{\tau'}L^{\phantom{\dagger}}_{\tau'}\!: + \,  L
\rightarrow R), \label{LCHee}
 \end{eqnarray}
with $\tau, \tau' = \pm$ summed over, and where $g_{1} \!\sim\! \tilde{V}(k\!\sim\!2k_F)$ and
$g_{2}\!\sim\! \tilde{V}(k\!\sim \!0)$, $\tilde{V}(k)$ being the Fourier transform of the screened Coulomb
potential $V(n-n')$ in Eq. (\ref{ee}). The backscattering $\sim g_1$ is weak in a semiconductor
structure and renormalizes to zero at
low energies also in the presence of spin-orbit interactions \cite{Schulz}. In effect we are thus left with
only the dispersive and forward scattering channels $\sim g_2$ in Eq. (\ref{LCHee}), to be added to
${\cal H}_1$ and ${\cal H}_2$ from Eqs. (\ref{linear-Hamiltonian}) and (\ref{2linear}). Passing to a bosonized formalism
\cite{Giamarchi_book_04}, the resulting full Hamiltonian density can be written as
${\cal H} = {\cal H}^{(1)} + {\cal H}^{(2)} + {\cal H}^{(12)}$ with
\begin{eqnarray}
{\cal H}^{(i)} \!&\!=\!& {\cal H}^{(i)}_0 \!+\!\!\frac{\lambda \,\delta_{i2}}{\sqrt{\pi K}a}\!\cos(\!\sqrt{4\pi K}\phi_2) \partial_x\theta_2, i\!=\!1,2 \label{bosoni} \\
{\cal H}^{(12)} \!&\!=\!& \frac{g_2K}{\pi}\partial_x \phi_1 \partial_x \phi_2 \label{boson12},
\end{eqnarray}
where $K \approx (1\!+\!g_2/\pi v_F)^{-1/2}$. Here ${\cal
H}^{(i)}_0\!=\! u[(\partial_x \theta_{i})^2 \!+\! (\partial_x
\phi_{i})^2]$ is a free boson theory with $u\! \approx \!v_F/2K$,
and with $\theta_i$ the dual field to $\phi_i$. The indices ``1''
and ``2'' tagged to the fields label the two branches originating from
Eqs. (\ref{linear-Hamiltonian}) and (\ref{2linear}).  

We should point out that our fields $\phi_i$ ($i\!=\!1,2$) are rotated with respect to the conventional bosonic fields $\phi^{R,L}_\tau$ ($\tau\!=\!\pm$) \cite{Shankar} representing the original fermion fields $R_{\tau}$ and $L_{\tau}$,  %
$\phi_i=\phi^R_\pm+\phi^L_\mp$, with upper (lower) sign attached to $i\!=\!1\ (i\!=\!2)$. This nonstandard spin-mixing basis $\{\phi_i\}$ is suitable for revealing how the non-diagonal term in Eq. (\ref{2linear}) combines with the e-e interaction in Eq. (\ref{LCHee}) to gap out the states near $\pm k_F \pm q_0$: The term in Eq. (\ref{2linear}) transforms into the sine-Gordon-like potential in Eq. (\ref{bosoni}) \cite{Malard2}, controlled by e-e interactions through the Luttinger liquid $K$-parameter. 
As we shall see, the theory brought on the form of Eqs. (\ref{bosoni}) and (\ref{boson12}) can be efficiently handled by using an adiabaticity argument.

To make progress we pass to a Lagrangian formalism by Legendre transforming Eqs. (\ref{bosoni}) and  (\ref{boson12}).
Using that $\Pi_i = \sqrt{K}\partial_x \theta_i$ serves as conjugate momentum to $\phi_i/\sqrt{K}$,
$\Pi_i$ can be integrated out from the partition function $Z$, with the result
\begin{equation}
Z \sim  \int {\cal D}\phi_1 {\cal D}\phi_2 e^{-(S^{(1)} + S^{(2)} + S^{(12)})},
\end{equation}
with Euclidean actions
\begin{eqnarray} \label{Actions}
\! \! \! \! \! \! \! \! S^{(i)} \!&\!=\!&\! S_0^{(i)} \! - \delta_{i2} \frac{m_0}{\pi a}\!\int\! d\tau dx \cos(\sqrt{16\pi K}\phi_2),  i\!=\!1,2   \label{branch1and2} \\
\! \! \! \! \! \! \! \! S^{(12)} \!&\!=\!&\! \frac{g_2 K}{\pi}\! \int \!d\tau dx \partial_x \phi_1 \partial_x \phi_2. \ \ \ \label{branch12}
\end{eqnarray}
Here $S_0^{(i)}= (1/2) \int d\tau dx [(1/v)(\partial_{\tau}\phi_i)^2 + v(\partial_x \phi_i)^2]$ is a
free action with $v=2u$, and $m_0 ={\lambda}^2 /4Kva$.

Having brought the theory on the form of Eqs. (\ref{branch1and2}) and (\ref{branch12}), valid for a Rashba modulation with
$Q=2(k_F+q_0)$, we first consider the auxiliary problem where the amplitude $g_2$ in Eq. (\ref{branch12}) is replaced
by a tunable parameter, $g_2'$ call it. Putting $g_2'=0$ and refermionizing $S^{(1)}$ we then obtain a helical Dirac action for the first branch (with Fermi points $\pm k_F \mp q_0$), with the second branch (with Fermi points $\pm k_F \pm q_0$) described by a sine-Gordon action, $S^{(2)}$. The cosine term in $S^{(2)}$ becomes RG relevant for $K<1/2$, driving this branch to a stable fixed point with massive soliton-antisoliton excitations \cite{Malard2}. The energy to create a soliton-antisoliton pair defines an insulating gap $\Delta$, and one finds from the exact solution of the sine-Gordon model \cite{Zamolodchikov} that
\begin{equation} \label{gap}
 \Delta = c(K)\Lambda(\frac{m_0}{\Lambda})^{1/(2-4K)}, \ \ K<\frac{1}{2},
\end{equation}
where $\Lambda =v/a$ is an energy cutoff, and $c(K)$ is
expressible in terms of products of Gamma
functions. The opening of a gap implies that the field $\phi_2$ gets
pinned at one of the minima of the cosine term. Thus, in the
neighborhood of the fixed point  its gradient is suppressed with the
effect that the action $S^{(12)}$ remains vanishingly small also
after $g_2'$ has been restored to its true value, $g_2' \rightarrow
g_2$. In particular, it follows that $S^{(12)}$ cannot close the
gap. Note that this ``argument by adiabaticity'' is perfectly well
controlled as the approach to a stable fixed point rules out any
non-analyticities in the spectrum. In summary, when $K<1/2$, a
Rashba modulation $Q=2(k_F+q_0)$ opens a gap in the second branch
which becomes insulating, {\em leaving behind a conducting helical
electron liquid} in the first branch (see FIG 2(a)).
\begin{figure}
\includegraphics[scale=0.33,angle=0]{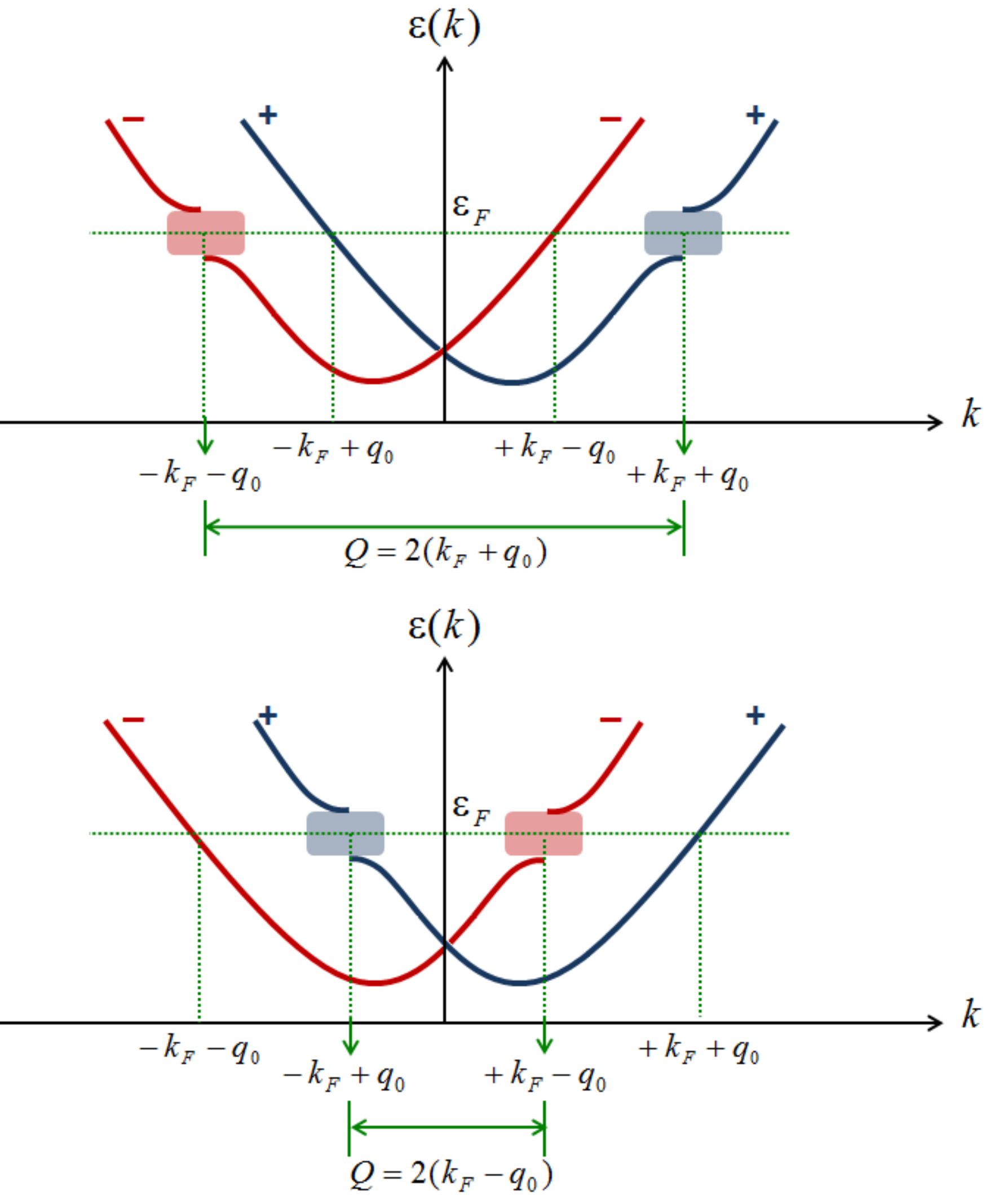}\\
\caption{(Color online) Schematic plot of the dispersion relations for
the two types of helical liquid phases, with (a)  $Q = 2(k_F + q_0)$ and (b) $Q = 2(k_F - q_0)$.} \label{Fig_Spectrum_3}
\end{figure}

The analysis above is readily adapted to the case with $Q=2(k_F-q_0)$, and one finds that the gap now opens in the first branch.
Note that our results remain valid in the presence of the weakened
commensurability condition $|Q-2(k_F + \tau q_0)| \ll {\cal O}(1/a) \, \mbox{mod} \,2\pi, \tau = \pm$, as this condition still allows us to throw away
the rapidly oscillating terms in the low-energy limit of $H_\text{R}^\text{mod}$.

Our interpretation of the dynamically generated gap $\Delta$ as an
effective band gap $-$ as in FIG. 2 $-$ draws on a result by
Schulz \cite{hjSchulz} where a bosonized theory similar to that
defined by our Eqs. (\ref{bosoni}) and (\ref{boson12}) 
is refermionized into a non-interacting two-band model, with the
bands separated by a gap corresponding to the dynamic
gap of the bosonized theory. This picture $-$ while heuristic only $-$
helps to conceptualize  the role of the commensurability conditions for the emergence
of the synthetic HL.

The fact that {\em e-e} interactions can open a gap in an
HL is well-known from the literature \cite{Wu,Xu,LutchynOreg}. In particular,
Xu and Moore \cite{Xu} noted that if a dynamically generated gap opens in one of two
coexisting Kramers' pairs ({\em alias} `branches' 1 and 2 in our model), this gives rise to a
stable HL in the other pair. Their observation pertains to the case where the scattering within
each branch is governed by distinct strengths of the {\em e-e} interaction: a gap
may then open in the branch with the stronger interaction. 
For this reason the Xu-Moore observation does not apply to the realistic case of
of a single quantum wire with the same interaction strength
in the two spin-split branches. This is where our proposal injects a
novel element into the picture: By properly combining a modulated Rashba spin-orbit interaction with a Dresselhaus
interaction we find that the gap-opening mechanism from {\em e-e} interactions can indeed be triggered in such a way as to
open a gap in one of the branches only, leaving behind a stable HL  in the other.
This HL is of a new type compared to the ones hitherto probed experimentally: It owes its existence neither to
being `holographic' \cite{JapanReview} (like the edge states of an HgTe QW \cite{Konig}) nor to being `quasi-helical' \cite{Braunecker}
(as is the case for magnetic-field assisted HLs \cite{Quay}). The time-reversal analogue of the notorious fermion-doubling problem
\cite{Nielsen} is instead circumvented by the fact that the gapped branch breaks time-reversal symmetry {\em spontaneously} by developing
a spin-density wave \cite{WuDiscussion}.  As there is no need to apply a magnetic field to realize the synthetic HL, it escapes the complications
from time-reversal symmetry breaking that mar a quasi-helical liquid \cite{Braunecker}. By this, it becomes an attractive candidate 
for renewed Majorana fermion searches \cite{Kouwenhoven}. 

Having established a proof of concept that a synthetic HL can be
sustained in a quantum wire by application of electric fields only,
is our proposal also a `deliverable' in the lab? The query can be
broken down into three specific questions: (i) Is it feasible to
realize a regime with sufficiently strong {\em e-e} interactions (as
required by the condition $K<1/2$)?  (ii) Can the size of the
gap $\Delta$ be made sufficiently large to block thermal
excitations at experimentally relevant temperatures? (iii) Is the
synthetic HL robust against disorder?

To answer these questions, we take as case study a quantum wire patterned in an InAs quantum well (QW) \cite{LiangGao,Giglberger}.
Starting with (i),
a detailed analysis yields that
\begin{equation}
 \tilde{V}(k \sim 0) \approx \frac{e^2}{\pi \epsilon_0 \epsilon_r} \ln(\frac{2d}{\eta}) + {\cal O}(\frac{\eta^2}{d^2})
\end{equation}
with $\eta$ the half width of the wire, and where
$\epsilon_r$ is the averaged relative permittivity of the dopant and
capping layers between the QW and a metallic back gate at a distance
$d$ from the wire  \cite{ByczukDietl}. The commonly used In$_{1-x}$Al$_{x}$As capping
layer has $\epsilon_r \!\approx\! 12$ when $x\!=\!0.25$, with roughly the
same value when doped with Si. With $\eta \approx 5$ nm and $v_F
\!\approx\! 6 \times 10^5$ m/s \cite{PB}, taking $d>1\, \mu$m and
using that $g_2 = 4 \tilde{V}(k\!\sim \!0)/\pi \hbar$
\cite{Giamarchi_book_04}, one verifies that $K \approx
(1\!+\!g_2/\pi v_F)^{-1/2}  <1/2$. Thus, the desired
``strong-coupling'' regime is attainable without difficulty.

Turning to (ii), we need to attach a number to the gap $\Delta$ in Eq. (\ref{gap}). Reading off data from Ref. \onlinecite{LiangGao}, applicable when the InAs QW is separated from the top gates by a solid PEO/LiClO$_4$  electrolyte, the Rashba coupling $\hbar \alpha$ is found to change from $0.4 \times 10^{-11}$ eVm to $2.8 \times 10^{-11}$ eVm when tuning a top gate from $0.3$ to $0.8$ V. With $a \approx 5$ \AA \  \cite{PB}, we may thus take $\hbar \gamma_{\text{R}} = 8$ meV and $\hbar \gamma^{\prime}_{\text{R}} = 60$ meV, assuming that [the spacers between] the top gates in Fig. 1 are biased at [0.3 V] 0.8 V. As for the Dresselhaus coupling, experimental data for InAs QWs come with large uncertainties. We here take $\hbar \gamma_{\text{D}} = 5 $ meV, guided by the prediction that $1.6 < \alpha/\beta  < 2.3$ in conventionally gated structures 
 \cite{Giglberger}. Inserting $\lambda = a \gamma^{\prime}_{\text{R}}\gamma_{\text{D}}(\gamma_{\text{R}}^{2}+\gamma^{2}_{\text{D}})^{-1/2}$ into Eq. (\ref{gap}), and choosing, say, $K=1/4$ with $c(1/4)= 1$ \cite{Zamolodchikov} we obtain $\Delta \approx 0.3$ meV (with smaller values of $K$ producing a larger gap).  While this value of $\Delta$ is much smaller than the bulk gap in an HgTe QW \cite{Konig}, it is still large enough $-$ with safe margins $-$ to protect the synthetic HL at sub-Kelvin temperatures. This allows to probe it by standard quantum transport experiments. It is here interesting to note that a recent proposal for an ``all-electric" topological insulator in an InAs double well arrives at an inverted band gap of roughly the same size as our interaction-assisted gap \cite{ErlingssonEgues}. 

Finally, let us address (iii). As shown in Refs. \onlinecite{Wu} and \onlinecite{Xu}, a 1D helical liquid may undergo a localization transition due to disorder-generated correlated two-particle backscattering. A case in point is when a Rashba interaction is present \cite{SJJ}, 
with a fluctuating component $\alpha_{\mbox{\footnotesize{rand}}}(x)$ from the random ion distribution in nearby doping layers \cite{Sherman}. Fortuitously, the localization length $\xi_{\text{rand}}$ for an InAs wire, making the usual assumption that $\sqrt{\langle \alpha^2_{\mbox{\footnotesize{rand}}}(x) \rangle} \approx \langle \alpha(x) \rangle$ \cite{Sherman}, turns out to be much larger than the renormalization scale $\xi = \hbar v/\Delta$ at which the helicity gap develops \cite{AS}. Moreover, estimates of the elastic mean free path $\ell_{\text{e}}$ for InAs quantum wires \cite{YYCC} show that $\xi < \ell_{\text{e}} < \xi_{\text{rand}}$ when $1/5 \!<\! K \!<\! 1/2$ and $\alpha_{\mbox{\footnotesize{rand}}}(x) \!<\! 4 \times 10^{-11}$ eVm. It follows that the synthetic HL is well protected within these parameter intervals.

In summary, we have unveiled a scheme for producing an interacting
helical electron liquid in a quantum wire using electric fields
only, exploiting an interplay between a Dresselhaus- and a spatially
periodic Rashba spin-orbit interaction. This synthetic helical
liquid is of a different type than existing
varieties, being neither `holographic'  \cite{Konig} nor
`quasi-helical'  \cite{Quay}. While a number of nontrivial
design criteria have to be satisfied for its realization in the
laboratory, none of them are beyond present-day capabilities.
Indeed, considering the principal simplicity and robustness of the
required setup, the synthetic helical liquid could become a
workhorse for exploring many of the intriguing phenomena associated
with helical electrons in one dimension.

\noindent We thank D. Grundler, K. Le Hur, and A. Str\"om  for valuable comments and suggestions.
This work was supported by SCOPES Grant IZ73Z0$_{-}128058/1$ (G.I.J.) and Swedish
Research Council Grant  621-2011-3942 (H.J.).

\end{document}